\documentclass[utf8]{FrontiersinHarvard} 

\usepackage{url,hyperref,microtype,subcaption}
\usepackage[onehalfspacing]{setspace}
\usepackage{natbib}
\usepackage{mathtools}
\usepackage{amsmath}
\usepackage{amsfonts}
\usepackage{amssymb}
\usepackage{graphicx}

\hypersetup{linkcolor=blue, citecolor=cyan, filecolor=cyan, urlcolor=blue, linktoc=all}

\def\keyFont{\fontsize{8}{11}\helveticabold }
\def\firstAuthorLast{Gamble {et~al.}} 
\def\Authors{Ronald Gamble\,$^{1,2,3,*}$, Jordan Forman\,$^{1}$, Amethyst Barnes\,$^{1}$, Gokul Srinivasaragavan\,$^{1,2}$, Isiah Holt\,$^{1,2}$ Marvin Jones Jr.\,$^{2,4}$}
\def\Address{$^{1}$NASA Goddard Space Flight Center, Greenbelt, MD, United States\\
$^{2}$University of Maryland-College Park, College Park, MD, United States\\
$^{3}$Center for Research and Exploration in Space Science \& Technology, University of Maryland-College Park, College Park, MD, United States\\
$^{4}$Indiana University, Bloomington, IN, United States}
\def\corrAuthor{Ronald Gamble}

\def\corrEmail{ronald.s.gamble@nasa.gov}

\begin{document}
\onecolumn
\firstpage{1}

\title {Multi-Messenger Emission Characteristics of Blazars} 

\author[\firstAuthorLast ]{\Authors} 
\address{} 
\correspondance{} 

\extraAuth{}

\maketitle

\begin{abstract}
Multi-Messenger observations and theory of astrophysical objects is fast becoming a critical research area in the astrophysics scientific community. In particular, point-like objects like that of BL Lac, flat spectrum radio quasars (FSRQ), and blazar candidates of uncertain type (BCU) are of distinct interest among those who look at the synchrotron, Compton, neutrino, and cosmic ray emissions sourced from compact objects. Notably, there is also much interest in the correlation between multi-frequency observations of blazars and neutrino surveys on source demographics. In this review we look at such multi-frequency and multi-physics correlations of the radio, X-ray, and $\gamma$-ray fluxes of different classes of blazars from a collection of survey catalogues. This multi-physics survey of blazars shows that there are characteristic cross-correlations in the spectra of blazars when considering their multi-frequency and multi-messenger emission. Accompanying this will be a review of cosmic ray and neutrino emissions from blazars and their characteristics.

\tiny
 \keyFont{ \section{Keywords:} High-Energy Astrophysics, Multi-Messenger Astrophysics, Supermassive Black Holes, Blazars, Active Galactic Nuclei, $\gamma$-ray, x-ray} 
\end{abstract}

\section{Introduction} \label{sec: intro}



Active Galactic Nuclei (AGN) are the largest, more luminous, persistent extragalactic objects observed in the Universe. These sources feature emission across the full gamut of electromagnetic spectra, from radio to $\gamma$-ray up to ultra high-energy cosmic rays. AGN, in general, encompass a large population of the high-energy $\gamma$-ray sources in the known Universe, comprising nearly 61.4\% of the 5064 $\gamma$-ray sources in the most recent completed update to the \textit{Fermi}-LAT 4FGL catalog \citep{Abdollahi2020FermiLA}. Blazars, and other point-like objects such as misaligned AGN or radio galaxies (e.g., \citealt{Abdo2010}) and Narrow-Line Seyfert 1 galaxies (\citealt{Dammando2019}), that feature similar emission patterns and mechanisms, play an essential role in our understanding of the high-energy Universe, potentially revealing crucial information about the evolutionary process of itself and the host galaxy. Blazars are of particular interest as they allow for direct observations of the relativistic jet emission and the resulting luminosity amplification due to the Doppler boosting of the emission. They are characterized by their extreme variability, high-polarization, radio-core dominance, and superluminal velocities \citep{2009ASPC..402..307L, 2016ApJS.226...20F}, and vary widely in timescales ranging from minutes to hours (intra-day variability), weeks to months (short-term variability), and months to years (long-term variability) \citep{1995ARA&A..33..163W, 2016MNRAS.462.1508G}. They are known to show two prominent broad-spectral features: the first peak is the result of synchrotron radiation and the second bump is potentially the result of inverse-Compton emission \citep{2016MNRAS.462.1508G, 2020ApJ...891..170V} that dominates leptonic models. The corresponding hadronic models in blazar Spectral energy distributions (SED) result from the higher-energy proton-synchrotron emission resulting from cascades of protons and pions in photo-meson productions \citep{bottcher2007, cerruti2020}. Blazars are categorized by two main subclasses: BL Lacertae objects (BL Lac), flat-spectrum radio quasars (FSRQ) \citep{galaxies10010035, Mohana_A_2023,Kramarenko_2021,2016ApJS.226...20F,2018Ap&SS.363..142Z}, along with a somehwat chameleon type of subclassification called \textit{changing-look blazars} \citep{Kang_2024}. The most notable differences between the two classes are the contrasts in emission lines. BL Lacs produce weakly peaked emission lines, while FSRQs produce very strong emission lines \citep{2009ASPC..402..307L}. The history of blazar unification has been a long-standing problem in AGN observations \citep{urry1995unified, rieger2019gamma,fossati1998unifying, padovani2017active}. 

The \textit{Fermi}-LAT collaboration \citep{Atwood_2009} has generated one of the most extensive catalogues of AGN in the high-energy regime \citep{Abdollahi2020, Ballet2023, Ajello2020}. A growing number of developing probe and mission concepts are dedicated to the multi-messenger aspects of observing these energetic objects with variable emission. Additionally, when considering correlations of higher energy observations with radio emissions of blazars, the joint MOJAVE-FERMI \citep{mojavefermi_cat} catalogue correlates these emission regimes observed by \textit{Fermi}-LAT and MOJAVE collaborations. Similarly on the lower end of the frequency spectrum, the MOJAVE (Monitoring Of Jets in Active galactic nuclei with VLBA Experiments) \citep{Lister2009} is stated as being a long-term program that observes the brightness and polarization of radio jets in AGN. Furthermore, there continues to be sources added to the joint MOJAVE-FERMI AGN catalogue \citep{Kramarenko_2021}. Recommendations from the Pathways to Discovery in Astronomy and Astrophysics for the 2020s (ASTRO2020) \citep{ASTRO2020} have generated a number products and initiatives that begin to prioritize science gaps for time-domain and multi-messenger (TDAMM) \citep{TDAMM_whitepaper} astrophysics. The $\gamma$-ray Transient Network Science Analysis Group (GTN SAG) \citep{burns2023gammaray} and various workshops and conferences that solicit community synergy like that of the TDAMM workshop: The Dynamic Universe: realizing the science potential of Time Domain and Multi-Messenger Astrophysics held as a result of the recommendations from \citet{ASTRO2020}.

The remainder of this review is organized as follows. In section \ref{blazar_physics} we provide a focused description of state of the art physical characteristics of blazars and their emitted jets across a multi-physics regime looking at the intersecting physics of jet launching. In section \ref{blazar_variability} we review current efforts that explore multi-spectral correlations and variability in blazars. Lastly we end this paper with a discussion on multi-messenger science gaps making parallels with other high-energy point-like objects that show similar emission characteristics as blazars. This section also highlights ongoing efforts and projects that attempt to reveal new areas of scientific interest in relation to a central black hole.

\section{Multi-Physics Characteristics of Blazars}\label{blazar_physics}
\subsection{Power Spectrum}
Relativistic jets comprise of the non-thermal emission within the AGN spectra, ranging from synchrotron sources of radio emission to higher energy $\gamma$-ray and even cosmic ray emissions. The power spectrum associated with synchrotron and self-synchrotron emission can be determined by using 
\begin{equation}\label{synchrotron_spec}
P(\nu) = \frac{\sqrt{3}e^{3}B\sin{\alpha}}{m_{e}c^{2}}\left(\frac{\nu}{\nu_{c}}\right)\int^{\infty}_{\nu/\nu_{c}}K_{5/3}(\eta)d\eta,
\end{equation}
where the critical frequency, $\nu_{c}$, is given by 
\begin{equation}
\nu_{c}=\frac{3}{2}\gamma^{2}\nu_{G}\sin{\alpha}
\end{equation}
, with $\nu_{G}$ as the gyrofrequency. The parameters $B, \alpha$, and $\nu$ are the magnetic field strength, pitch angle, and emission frequency, respectively. While the integral in the synchrotron power function here is characterized by the modified Bessel function of the second kind $K_{5/3}(\eta)$, where $\eta$ is defined as a ratio of the frequency to critical frequency $\nu_{c}$.
Additionally, their spectra can be determined using various observational data analysis methods and SED correlation schemes \citep{Homan_2021}. Current data analysis from observational missions have shown that the SEDs of BL Lacs and FSRQs exhibit significant continuum variability in their observed frequency bands \citep{Abdollahi2020FermiLA,2020ApJ...891..170V, Harris2006,Mohana_A_2023}. This spectral data can be connected back to the black hole-disk system to infer local properties of the surrounding accretion disk (i.e. matter content, dust/plasma temperature, particle accelerations/scatterings, etc.), but is limited in describing the gravitationally induced dynamics of the relativistic jet \citep{gamble2022spin}.
\begin{figure}[h]
\centering
\includegraphics[scale=0.4]{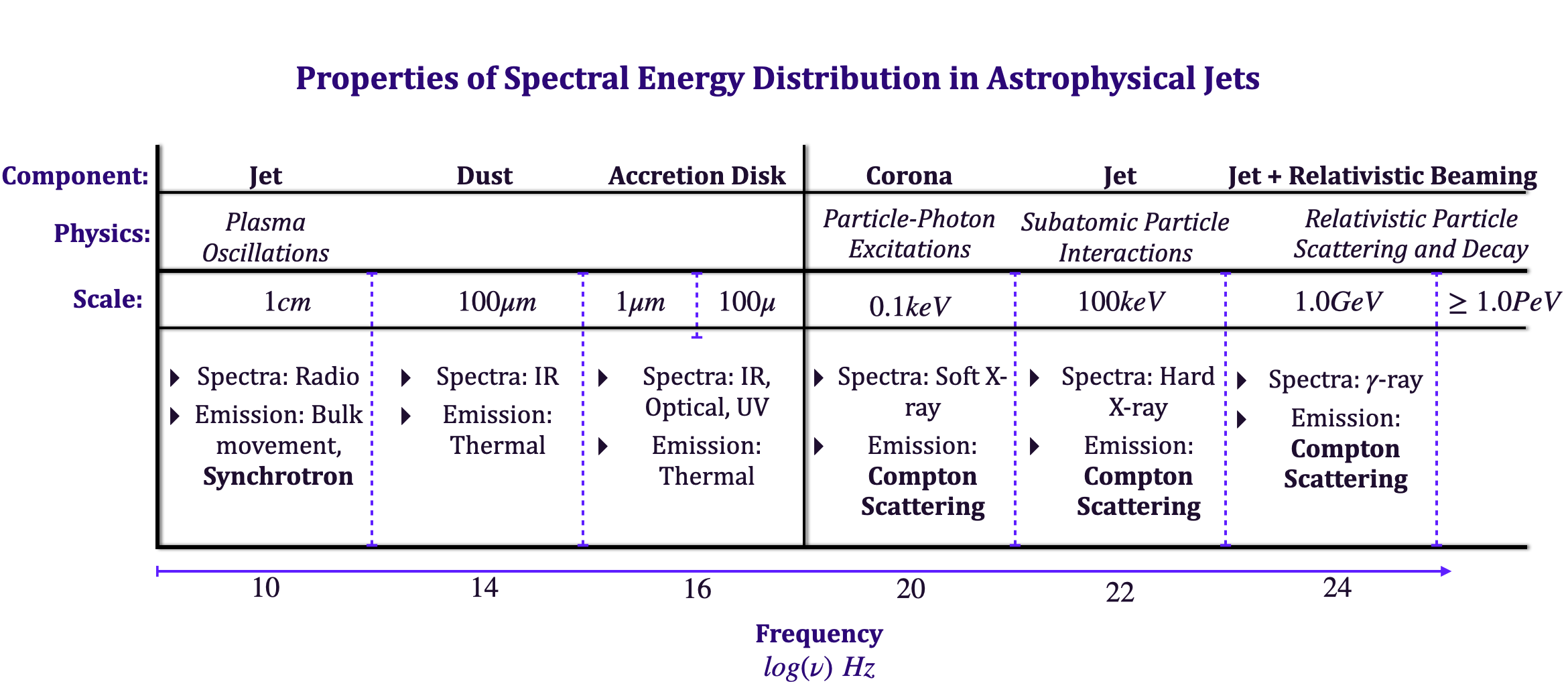}
\caption{Properties of relativistic jet spectra and their corresponding radiation transfer phenomena. \citep{gamble2022spin}}
\end{figure}

\subsection{Jet Emission Mechanisms}
Currently, the mechanisms for relativistic jet emission associated with AGN, and other high-energy astrophysical objects like $\gamma$-ray bursts (GRB), and microquasars, has been of current interest in the astrophysics scientific community. Jet formation theory and emission is a major problem yet to be solved in high-energy astrophysics. One of the most widely argued models for describing this type of emission has been the Blandford-Znajek (BZ) process \citep{bz77}. This process describes the rotational energy extraction from black holes involving the torsion of magnetic field lines resulting in Poynting flux dominated outflows parallel to the rotation axis of the central object \citep{bz77} and \citep{znajek77}
\begin{equation}\label{bzlumn}
L_{BZ} = f(\alpha_{H}) B_{\phi}^{2} r^{2}_{s} c 8\pi^{-1}.
\end{equation}
Where equation \ref{bzlumn} is the BZ-luminosity. Here we define the parameters $\alpha_{H}, B_{\phi},r_{s}$ as the spin parameter of the black hole horizon, magnetic field strength in the $\phi$-direction, and the corresponding Schwarzschild radius, respectively.
The nature of such highly complex energetic emission mechanisms from these systems, which feature event horizons in rotating spacetimes, has been studied extensively over the last few decades \citep{Williams2004, Williams1995,gamble2022spin,altgr_bz,bzcurrent1,king_2021}. Recent numerical and observational models incorporating magnetohydrodynamic (MHD) and general relativistic magnetohydrodynamic (GRMHD) methods have shown that a major contribution to jet outflows are from the poloidal magnetic field configurations from relativistic matter accreting on to the central object \citep{Komissarov_2005, Contopoulos2014,koide2020,EHT:2022urf}.
Unanswered questions on the relativistic nature of these jets involve figuring out how particles that make up the jet's content are accelerated to ultra-relativistic speeds whose Lorentz factors are $\Gamma_{Lorentz} >10$. What is the origin of the relativistic particles that produce non-thermal radiation we observe? And how do these jets become \textit{matter loaded}? Focusing on the theoretical aspects of jet formation mechanisms, there are still fundamental questions that continue to remain unresolved. One of which is the causal connection of the jet to the exterior Kerr spacetime. An application of the BZ-process to alternative or extensions of general relativity by \cite{altgr_bz} has shown the versatility in the decades-old theory but, again, exhibits how the BZ-process needs extensions to incorporate the sources of the magnetic fields it describes \citep{king_2021, bzcurrent2}.

As we have seen a relativistic jet described as a beam of light carries linear momentum, and thus is influenced by an appreciable amount of external angular momentum in both the non-relativistic and relativistic regimes. This angular momentum would then be dependent on the origin of an associated coordinate system, owing to the intrinsic gauge dependence of angular momentum in fundamental physics descriptions. If we then proceed to describe BL Lac and FSRQ blazars as energetic point-sources, we can infer the physical characteristics of the jet emission as relativistic beams transported across galactic distances. These point sources should then inherently carry a rotational symmetry corresponding to rotated field lines with respect to the host black hole \citep{gamble2022spin}. The following equations of motion, specifically under the influence of curved spacetime near the jet launching region, illustrate the complexities of jet launching from the supermassive black holes of blazar types. Here the potentials parameterizing particle paths in this near-horizon region are defined giving the set of Hamilton-Jacobi equations for each direction. It is easy to see the expected symmetries in the particle paths for the $t$ and $\phi$ directions. Here the functions $R(r)$ and $V(\theta)$ correspond to the traditional motions in the $r$ and $\theta$ directions.
\begin{subequations}
\begin{eqnarray}
\Sigma \frac{dr}{d\lambda} &=& \pm \sqrt{R(r)}\\
\Sigma \frac{d\theta}{d\lambda} &=& \pm\sqrt{V_{\theta}(\theta)}\\
\Sigma \frac{d\phi}{d\lambda} &=& -(\alpha_{H} E-L/sin^{2}\theta) + \alpha_{H} T/\Delta\\
\Sigma \frac{dt}{d\lambda} &=& -\alpha_{H}(\alpha_{H} Esin^{2}\theta - L)+(r^{2}+\alpha^{2}_{H})T/\Delta
\end{eqnarray}
\end{subequations}
where the functions, $T$, $R(r)$, and $V_{\theta}(\theta)$ are defined as
\begin{eqnarray}
T &\equiv& E(r^{2}+\alpha^{2}_{H})-\alpha_{H} L\\
R(r) &\equiv& T^{2} -\Delta\left[m_{0}^{2}r^{2}+(L-\alpha_{H} E)^{2} +Q\right]\\
V_{\theta}(\theta) &\equiv& Q - cos^{2}\theta\left[\alpha^{2}_{H}(m_{0}^{2}-E^{2})\frac{L^{2}}{sin^{2}\theta}\right]
\end{eqnarray}
Here $E$ and $L$ are the particle energy and angular momentum, respectfully, while $m_{0}$ is the rest mass of a test particle and $Q$ is identified as Carter's constant. 
The functions $\Sigma = r^{2}+\alpha^{2}_{H}cos^{2}\theta$ and $\Delta = r^{2} -Mr +\alpha^{2}_{H}$ are defined from the components of the Kerr spacetime for a rotating black hole of arbitrary mass. Within the context of this discussion on blazar jet emission it is then logical to think about, not only the particle distributions in jets, but also the intrinsic geometry of particle paths moving at high Lorentz factors above $\Gamma_{bulk}\geq 10-10^2$. Additionally, there have been efforts to incorporate nonequitorial instabilities that contribute to the $e^{-}/e^{+}$ pair production at $\gamma$-ray energies $\geq GeV$ around high spin $\alpha_{H} \geq 0.8$ black hole horizons in a description of jet launching \citep{Williams1995, Williams2004}.
Thus, removing some of the mystery of the physical mechanisms that causes some jets to twist and carry a proportionate amount of angular momentum from the black hole. It is then intuitive to think about how one can infer the mechanisms causing such polarization in the observed spectra. Observations of blazars and radio-loud AGN have shown that polarization states exist in the spectra from these sources \citep{cp_blazar2021,Homan_2021}.

\section{Multi-Spectral Variability of Blazars}\label{blazar_variability}

\subsection{Variability and Flaring of VLBI-Selected Blazars}
Observing the variability of blazars can reveal the necessary information to infer the composition of the jet emissions, the mechanisms behind the jet formation, changes in the accretion rate of the accretion disk, and can allow for localization of the innermost emitting regions \citep{2020ApJ...891..170V, 2016ASPC..505..107L}.  As the central SMBH at the cores of blazars accretes matter and forms the surrounding accretion disk, it launches relativistic jets which emit across the electromagnetic spectrum (radio to $\gamma$-rays) \citep{2016MNRAS.462.1508G}. 
\begin{figure}[h!]
\centering
\includegraphics[width=0.5\textwidth]{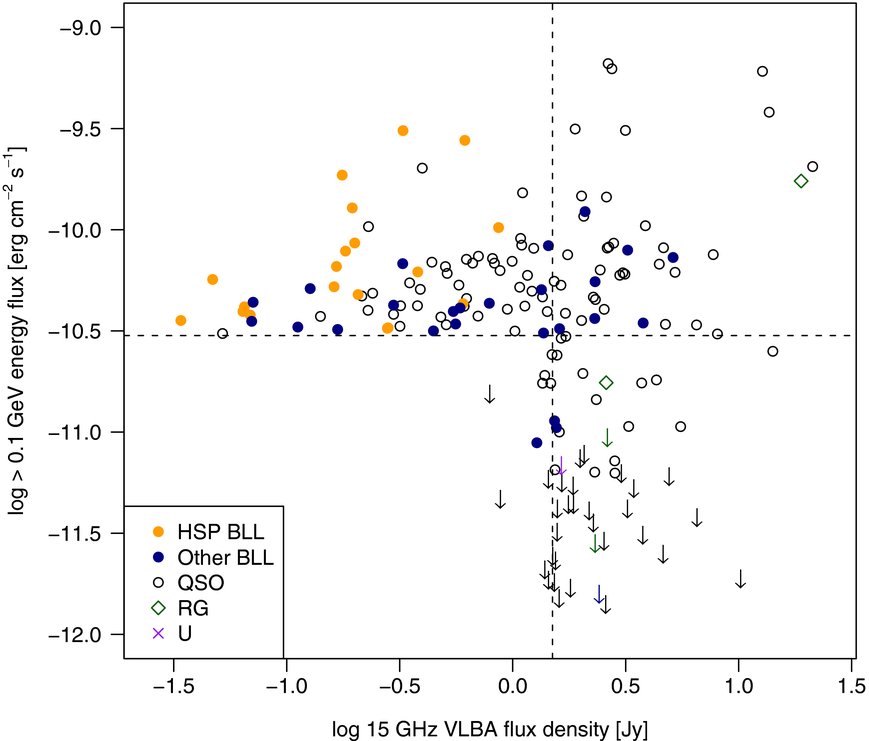}
\caption{Plot of 11-month Fermi average $>0.1 GeV$ energy flux vs. 15 GHz VLBA flux density of the joint blazar samples in \citep{mojavefermi_cat}. The filled circles represent BL Lac objects, with the HSP ones in orange and others in blue. The open circles represent quasars, the green diamonds radio galaxies, and the purple crosses optically unidentified objects. Upper limits on the $\gamma$-ray fluxes are indicated by arrows. All of the BL Lac objects are detected by the LAT, with the exception of J0006-0623. The vertical dashed line indicates the sample radio limit of $1.5 Jy$, and the horizontal dashed line indicates the $\gamma$-ray limit of $3 \times 10^{-11}erg$ $cm^{-2}$ $s^{-1}$. Figure and caption are sourced from the MOJAVE-\textit{FERMI}-LAT 1FGL catalogue \citep{mojavefermi_cat}.}\label{fig:mojave-fermi}
\end{figure}
Figure \ref{fig:mojave-fermi} shows such a distribution in the GeV energy flux associated with $\gamma$-ray emissions versus the VLBA flux for these radio-gamma correlated sources. This distribution shows differentiation between high synchrotron peak (HSP) BL Lacs that feature peaks in the range $\nu > 10^{15}$Hz and low synchrotron peaked (LSP) BL Lacs who fall in the range $\nu < 10^{14}$Hz \citep{sahakyan2020}. See discussions in \citep{Abdo_2010, giommi1994} for more detailed descriptions comparing HSP and LSP signatures for BL Lacs. 

\begin{figure}[h!]
\centering
\includegraphics[width=0.5\textwidth]{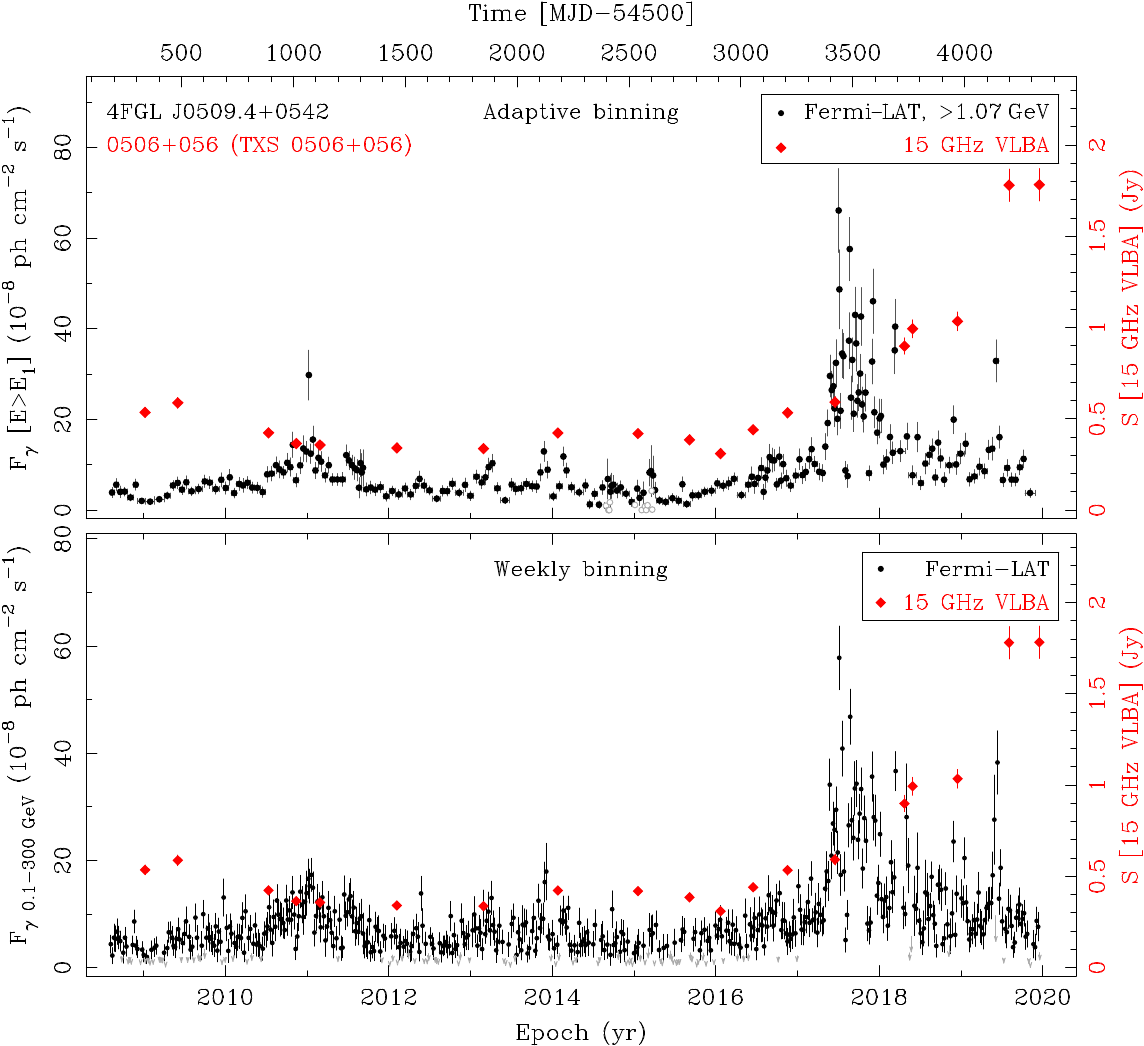}
\caption{Intra-week variability overlap of the $\gamma$-ray emission at $E_{ph}>1.07$ GeV (in black) and the VLBA radio emission at 15 GHz (in red) of the blazar TXS 0506+056 (4FGL J0509.4+0542). The two panels show the light curve with adaptive binning on the top, and with weekly binning on the bottom where the epoch spans 12 years. Multi-epoch light curve is sourced from the MOJAVE-\textit{FERMI}-LAT 1FGL catalogue \citep{mojavefermi_cat}.}\label{fig:TXS0506}
\end{figure}

In figure \ref{fig:TXS0506} below we can see that there exists a delayed variability in the radio emission for the blazar TXS 0505+056 (4FGL J0509.4+0542) as compared to its higher energy counterpart in the light curve at $E_{ph}>1.07$ GeV. This light curve, along with blazars in the MOJAVE-FERMI catalogue features this type of variability where the radio and $\gamma$-ray emission are correlated according to a respective time-lag.
There exists significant correspondence with the $\gamma$-ray flaring of TXS 0506+056 (4FGL J0509.4+0542) with neutrino incidence in the direction of  this blazar \citep{icecube.txs0506.2018}. Looking at the photo-meson production for HSP as state above, such particle interactions within the jets of highly energetic sources like TXS 0506+056 (4FGL J0509.4+0542) and PKS 0735+178 (4FGL J0738.1 + 1742) \citep{prince2023} is a testament  of the dynamic multi-messenger and multi-physics aspect of sources that feature  extremely accelerated ejecta. The correlation between the radio and very high-energy (VHE) $\gamma$-ray emission is a curious notion highlighting the new frontier of multi-messenger astrophysics in the modern era of astronomy. Additionally, HSP blazars with similar flaring characteristics are also likely to exhibit particle cascade mechanisms that produce cosmic-rays (high-energy nucleons and charged particles). The 116 sources in the MOJAVE-\textit{FERMI}-LAT 1FGL catalogue are a prototypical example of the type of variability blazars exhibit across multiple spectral frequencies. Note that the catalogue only correlates VLBI-selected 15 GHz radio-loud sources with significant correlation to their $\gamma$-ray peaks. The catalogue is sourced from \citet{Kramarenko_2021}, A decade of joint MOJAVE-Fermi AGN monitoring: localization of the $\gamma$-ray emission region, that features 331 sources with down selection to N-Blazars with significantly strong radio emission $(>= ~80\%)$, of the 331 total catalogue of sources. Both blazar classes have been reported to present strong correlations between the radio and $\gamma$-ray emissions (e.g, \citep{2014MNRAS.445..428M, 2016ApJS.226...20F, 2015MNRAS.450.2658M}, thus indicating that the production of these jet emissions coincides with a common mechanism.
\begin{figure}[h!]
\centering
\includegraphics[width=0.9\textwidth]{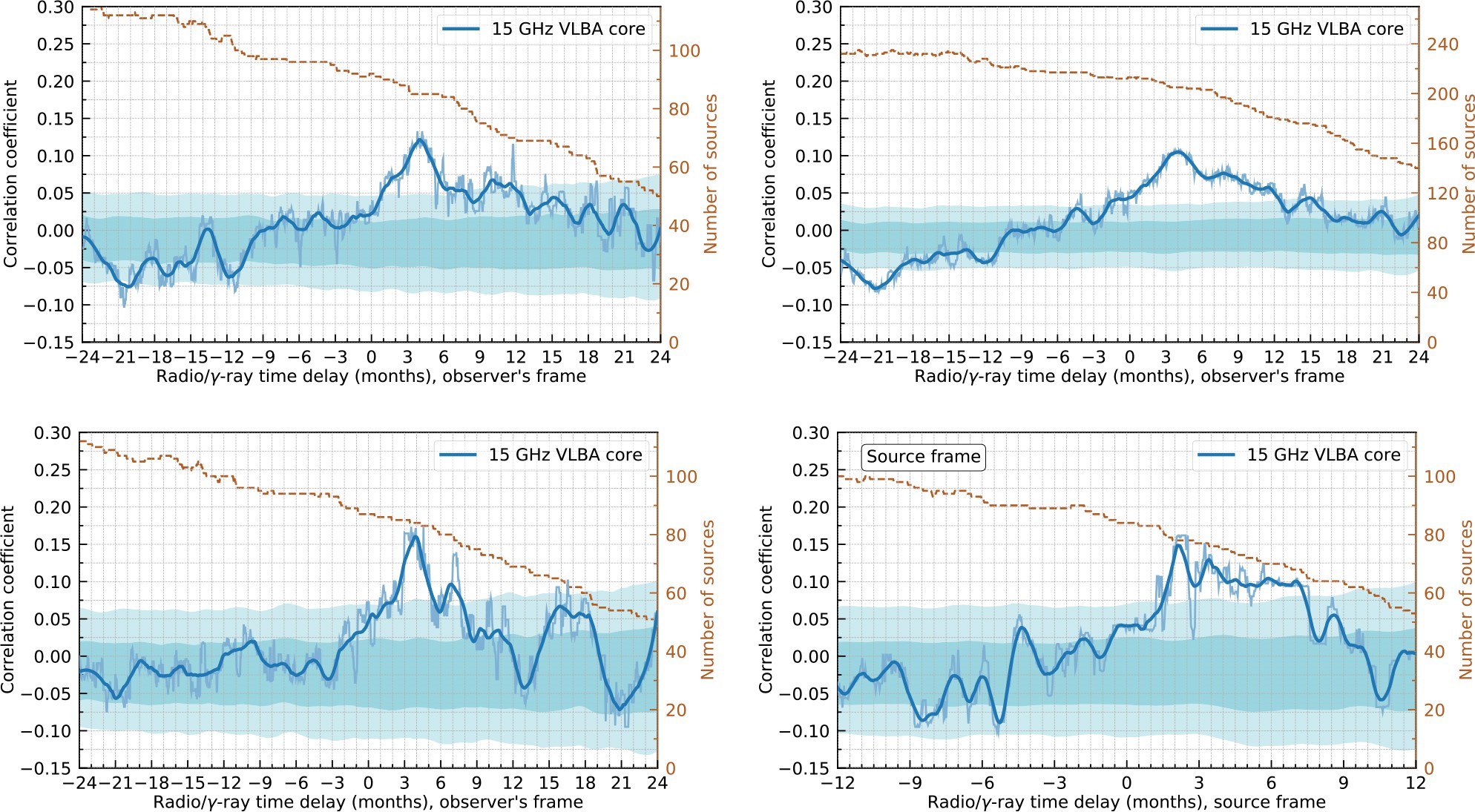}
\caption{Intra-week variability cross-correlation at 15GHz VLBA. Reproduced with permission from  \citep{Kramarenko_2021}}\label{fig:intra-week}
\end{figure}
A more extensive overview of radio VLBI/$\gamma$-ray catalogues of Blazars: MOJAVE-\textit{FERMI}-LAT 1FGL, National Radio Astronomy Observatory (NRAO) catalogues, Atacama Large Millimeter/submillimeter Array (ALMA), Event Horizon Telescope results and simulations, will be provided in subsequent papers looking at more details of cross-correlation in blazars.
Figure \ref{fig:intra-week} shows such intra-week variability at 15GHz in the time-domain. This variability illustrates the need for time-domain follow-up for energetic sources. We can see that on a month-to-month timescale the correlation strength peaks at $\sim 5$ months. This suggests that there could be significant observing campaign for follow-up observations. In regards to a multi-physics perspective, improved time-dependant theoretical models and GRMHD simulations are needed to decipher such physics.

\section{Discussion}\label{discussion}
\subsection{Blazar Parallels With $\gamma$-ray Bursts}
 Given the nature of the high-energy emission characteristics of BL Lac and FSRQ blazars, it is additionally safe to make the comparison to GRBs. Both types of high-energy sources are considered to be sourced by compact objects (i.e. SMBH, X-ray binaries, neutron star mergers, core-collapse supernovae, and stellar mass black holes). Both energetic phenomena exhibit similar physical characteristics when considering their respective ejecta mechanisms. It is of no coincidence that GRB and blazar jets also feature similarities in the spectral peaks illuminating commonalities in their respective radiation physics \citep{nemmen2012}. A more detailed description of these physical comparisons can be found in work highlighting such comparisons \citep{Lyu_2014,srinivasaragavan2023characterizing}. An even more interesting recent inclusion in the "AGN zoo" are \textit{changing-look blazars}. These are blazars that feature changes in their accretion processes, intrinsically changing from FSRQ-type to BL Lac and vise-versa \citep{Kang_2024}. This suggests that further investments in TDAMM science and its technological developments are needed to further elucidate dynamical properties of AGN with blazar types, BL Lac, FSRQ, and BCU.

\subsection{Ground-Based Follow-up}
\subsubsection{ALMA: Radio}
Specifically within the radio frequency regime, the ground-based ALMA \citep{alma} is extraordinary for observing, in general, AGN of different classifications, as it gives a perspective of these high-energy objects in the radio and infrared spectrum. With its ground-breaking interferometric array of 66 high-precision antennas, its performance results in high-resolution images with the brightness sensitivity of a single-antenna array \citep{alma1}. LSP BL Lac objects offer a distinctive spectral climb when comparing their $\gamma$-ray peak to their maximal synchrotron peaks \citep{Mohana_A_2023}, with blazars of type FSRQ almost exclusivley falling under LSP \citep{sahakyan2020}. Conversely, when analyzing the spectral correlation of HSP BL Lac objects with similar $\gamma$-ray energies the correlation is not as strong.$(<10GeV)$

Quasar PKS 1549-79 was previously observed by \cite{pks} in order to analyze its radio jet, making millimeter and very long baseline interferometry 2.3-GHz continuum observations. PKS 1549-79 is known as a radio-loud quasar, having a stronger radio emission and higher energy than the more common radio-quiet quasar (\cite{radioql}). PKS 1549-79 is also the closest one that has been caught merging with an AGN in the first phases of its evolution. \cite{pks} and their team also presented CO(1-0) and CO(3-2) observations of its molecular gas. Their results found that the massive outflow of 650 $M_\odot$ $yr^{-1}$  being confined to $r < 120$ pc of the inner galaxy suggests that the AGN is driving this outflow. The radio-quiet quasar SDSS J0924+0219 was observed by \cite{jo924} using 45 of ALMA’s antennas and the Very Large Array (VLA). It is evident that analyzing both, LSP and HSP blazars, contribute to a more compounded description of blazar models when looking at the entire non-thermal spectra of blazars in the AGN zoo. 

\subsubsection{IceCube: Neutrinos \& Cosmic-Rays}
The flaring and variability of the blazar spectra listed in the MOJAVE-FERMI catalogue, the \textit{Fermi}-LAT catalogues, and various others that feature high-energy $\gamma$-ray emission from blazars residing in their active phases are an important aspect to identifying the neutrino production from such sources (e.g. TXS 0506+056 (4FGL J0509.4+0542) and PKS 0735+178 (4FGL J0738.1 + 1742)). Looking closer at the particle production mechanisms, we can see that particle phenomenology associated with the electromagnetic and cosmic-ray producing interactions has overlap with their decay mechanisms as well. The photo-meson particle production in the accelerated environments of jets can be seen as illustrated in equations \ref{eqn:proton} and \ref{eqn:pion}, where protons scatter off of photons to produce a cascade of charged and neutral pions $(\pi^{0}, \pi^{+},\pi^{-})$.
\begin{eqnarray}\label{eqn:proton}
&&p + \gamma \rightarrow p' + \pi^{0}\\ \nonumber
&&p + \gamma \rightarrow n + \pi^{+}\\ 
&&p + \gamma \rightarrow p' + \pi^{+} + \pi^{-} \nonumber
\end{eqnarray}
This interaction of accelerated protons with $\gamma$-ray photons provides a precursor to the neutral and charge pions. The subsequent decay of $(\pi^{-}, \pi^{+})$ into a cascade of muons $(\mu^{+},\mu^{-})$ and neutrinos $(\nu_{e},\nu_{\mu})$ (of $e^{-}$ and $\mu^{+}$ types) and their respective symmetric (antimatter) pairs introduces the weak interaction into hadronic/meson blazar jet models.
\begin{eqnarray}\label{eqn:pion}
&&\pi^{0} \rightarrow 2\gamma\\ \nonumber
&&\pi^{+} \rightarrow \mu^{+} + \nu_{\mu} \rightarrow e^{+}+\nu_{e} + \overline{\nu}_{\mu} + \nu_{\mu}\\ 
&&\pi^{-} \rightarrow \mu^{-} + \overline{\nu}_{\mu} \rightarrow e^{-}+\overline{\nu}_{e} + \nu_{\mu} + \overline{\nu}_{\mu} \nonumber
\end{eqnarray}
Ultimately, the presence of these cascades detected by neutrino and Cerenkov telescopes is a prominent clue for finding relativistic protons in the jet \citep{cerruti2020, muecke1999photomeson}. The IceCube Neutrino Observatory \citep{icecube2017} has had significant progress in detecting neutrinos of astrophysical origin emanating from blazars. Blazars, such as TXS 0506+056 (4FGL J0509.4+0542) and PKS 0735+178 (4FGL J0738.1 + 1742), have been extensively studied in recent years \citep{icecube.txs0506.2018,padovani2015,prince2023}. Multi-messenger observations and their follow-up has thus proven to be a powerful methodology for determining the VHE characteristics of blazars. 
\section{Conclusions}
In this focused review of blazars of type FSRQ, BL Lac, and BCUs we have seen just how dynamic these point-like objects are regarding their relativistic properties. The multi-physical nature of such astronomical objects suggests that there are significant gaps in our understanding of their multi-messenger characteristics. The recommendations from the ASTRO2020 decadal survey offer an initiation of thoughts surrounding TDAMM science gaps. Further investments from the broader astronomy/astrophysics community are required to elucidate and decipher the true nature of blazars, their relativistic jet emission, and future multi-spectral analysis and missions. The utilization of unconventional thoughts and methodologies would prove useful in our quest to understand the energetic Universe. The synergy between radio (ALMA, MOJAVE, etc.), X-ray (IXPE, XRISM, Chandra, SWIFT, etc.), $\gamma$-ray (VERITAS, \textit{Fermi}-LAT, MAGIC, H.E.S.S.), and cosmic-ray/neutrino (IceCube) observations plays an important role in the analysis and theoretical modeling of variable energetic blazars, as it allows for more detailed observations of these objects.

\section*{Acknowledgements}
The material is based upon work supported by NASA under award number 80GSFC21M0002. This research has made use of data from the MOJAVE database that is maintained by the MOJAVE team \citep{mojavefermi_cat}. The authors wish to thank the reviewers for their valuable remarks and comments.
\bibliographystyle{Frontiers-Harvard.bst}
\bibliography{REF_full.bib}
\end{document}